\documentclass[preprint,12pt]{elsarticle}

\usepackage{amssymb}

\usepackage{newtxtext,newtxmath}
\usepackage{hyperref}

\usepackage[T1]{fontenc}
\usepackage{ae,aecompl}

\usepackage{graphicx}	
\usepackage{amsmath}	
\usepackage{amssymb}	

\journal{Astroparticle Physics}

\begin{document}

\begin{frontmatter}



\title{Unveiling the origin of  the gamma-ray emission in  \textsc{NGC 1068} with the Cherenkov Telescope Array}


\author{
A. Lamastra,$^{1,2}$
F. Tavecchio,$^3$
P. Romano,$^3$ 
M. Landoni,$^3$ 
S. Vercellone$^3$
\\
}

\address{
$^1$ INAF - Osservatorio Astronomico di Roma, via di Frascati 33, 00078 Monte Porzio Catone, Italy\\
$^2$ SSDC--ASI, via del Politecnico, 00133 Roma, Italy \\       
$^3$ INAF - Osservatorio Astronomico di Brera, via  E.\ Bianchi 46, I-23807, Merate, Italy
}

\begin{abstract}
Several observations are revealing the widespread occurrence of mildly relativistic wide-angle AGN winds strongly interacting with the gas of their host galaxy. 
   Such winds are potential cosmic-ray accelerators, as supported by gamma-ray observations of the nearby Seyfert galaxy \textsc{NGC 1068}  with the \textit{Fermi} gamma-ray space telescope.
   The non-thermal emission produced by relativistic particles accelerated by the  AGN-driven  wind observed in the circum-nuclear molecular disk of such galaxy is invoked to produce the gamma-ray spectrum. The AGN wind model predicts a hard spectrum  that extend in the very high energy  band which differs significantly from those corresponding to other models discussed in the literature, like starburst or AGN jet. We present dedicated simulations of observations through the Cherenkov Telescope Array (CTA), the next-generation ground based gamma-ray observatory,  of the very high energy spectrum of the Seyfert galaxy \textsc{NGC 1068} assuming the AGN wind and the AGN jet models. We demonstrate that, considering 50 hours of observations, CTA can be effectively used to constrain the two different emission models, providing important insight into the physics governing the acceleration of particles in non-relativistic AGN-driven outflows. This analysis strongly motivates observations of Seyfert and starburst galaxies with CTA in order to test source population  models  of the extragalactic gamma-ray and neutrino  backgrounds.

\end{abstract}


\end{frontmatter}



\section{Introduction}

Galactic winds driven by star formation and active galactic nuclei (AGN) play a key role in the evolution of galaxies. The galactic winds, powered by the momentum and energy  injected by massive stars in the form of supernovae (SN), are believed to regulate the rate of star formation in low mass galaxies. In high-mass galaxies,  galaxy formation models require AGN to inject energy and momentum into the surrounding gas in order to reproduce key observables of galaxy population, like the  bright end of the galaxy luminosity function, and the tight scaling relations between  supermassive black hole (SMBH) masses and galaxy bulge properties \citep[e.g.][]{King032,DiMatteo05,Croton06,Menci06,Somerville08,Lamastra10}.

One way of coupling AGN energy and momentum  with the gas in the host galaxy is through AGN-driven winds and jets.
Jets are highly collimated relativistic outflows of particles which characterize radio loud AGN (which represent  $\sim$ 10\% of the AGN population) and are especially prominent in blazars, in which one of the jets points close to our line of sight. 
The bulk of the AGN population does not show jet-like structure \citep[see][for a review]{Padovani17}. The active nucleus of these galaxies  is observed to  eject  wider-angle wind of lower velocity.
These  winds are observed in different ionization states (ionized, neutral atomic, and molecular) and at different spatial scales (from nuclear to galactic scales) through spectroscopy in the X-ray, UV, and NIR  bands, and  interferometric observation in the sub-millimitre band \citep[see][and references therein]{Fiore17}.

The most striking difference between jetted and non-jetted AGN is in the origin and  nature of their electromagnetic emission.  
As inferred from blazars, relativistic jets emit a large fraction of their energy  in the radio and gamma-ray bands through  non-thermal processes. Conversely, the  emission of non-jetted AGN is dominated by thermal emission in the UV-optical band produced by the accretion disk around SMBH. \\
Non-jetted AGN may emit also non-thermal radiation in the gamma-ray band, as indicated by the  detection of nearby Seyfert galaxies \textsc{NGC 1068}, \textsc{NGC 4945}, and the Circinus galaxy with the \textit{Fermi} gamma-ray space telescope \citep{Ackermann12,Hayashida13}.
These observations  has lead to consider non-jetted AGN as potential cosmic-ray (CR) accelerators, and it has also been suggested that  the diffuse extragalactic gamma-ray  background (EGB) 
measured by  the Large Area Telescope (LAT,  \citep{Atwood09} )  on board the  {\it Fermi} telescope \citep{Ackermann15} could be reproduced by the cumulative gamma-ray emission from the whole AGN population  (\citep{Wang16_gamma,Lamastra17}, but see \citep{Liu18}).

In this paper, we assess the potentiality of the Cherenkov Telescope Array (CTA)  to test  CR  acceleration models in non-blazar AGN, focusing on the test case of the prototypical Seyfert galaxy \textsc{NGC 1068} that is a nearby (D=14.4 Mpc) Seyfert 2 galaxy in which AGN-driven molecular wind, and  radio structures, similar to collimated outflows but weaker and slower than the jets  observed in blazars, have been  observed \citep{Krips11,Garcia14,Gallimore96,Wynn85,Gallimore06, Sajina11,Honig08}. 

Moreover, it is the brightest of the star forming galaxies detected   by \textit{Fermi}-LAT \citep{Ackermann12}.
The \textit{Fermi} collaboration interpreted the gamma-ray detection in terms of CR  associated to star-formation processes, adopting the paradigm that Galactic CR are relativistic particles  accelerated in SN-driven winds,  and that the gamma-ray emission is due to CR interaction with the galactic interstellar medium (ISM).   However,  some models assuming that the gamma-ray emission is entirely due to starburst activity underestimated the  \textit{Fermi}-LAT spectrum \citep{Yoast14,Eichmann15}.
This suggests that the active nucleus also contributes to the gamma-ray emission.  In this framework, two different theoretical models have been proposed so far.
A lepto-hadronic model in which the gamma-ray emission is  produced by electrons and protons accelerated by the shocks produced by the AGN-driven  molecular wind \citep{Lamastra16}, and a  leptonic model in which the gamma-ray emission is produced through inverse Compton (IC) scattering of low energy photons from the  relativistic electrons accelerated in the radio jet \citep{Lenain10}.

The gamma-ray spectra predicted by the AGN wind and AGN jet models differ significantly at energies $E\gtrsim$100 GeV, where imaging atmospheric  Cherenkov telescopes (IACT) are more sensitive than \textit{Fermi}-LAT. However, the  sensitivity of current IACTs does not allows us to test the existence of the emission in the very high energy (VHE)  band predicted by the AGN wind model. The construction of the CTA is expected to provide a broad (20 GeV-300 TeV) energy range, and  an average differential sensitivity a factor 5-20 better with respect to the current IACTs. We performed dedicated simulations of observations through  CTA of the gamma-ray spectrum of \textsc{NGC 1068} predicted by the  AGN wind and AGN jet models, in order to constrain the two competing scenarios.

The paper is organised as follows. In Section \ref{AGN_wind} we recall the basic points of  the AGN wind and AGN jet models and illustrate the predicted  spectral energy distributions (SED) used for the simulations; in Section \ref{lama_simulations} we describe the simulation set-up and present our results, while we give our discussion and conclusions  in Section \ref{discussion}.

\section{Theoretical  models}\label{AGN_wind}

In this section  we briefly describe the AGN wind and AGN jet models, and we compare the predicted gamma-ray spectra with  \textit{Fermi}-LAT data.

 \begin{table*} 	
 \caption{AGN  models and corresponding key quantities. W1$\div$W4= AGN wind models. Jet= AGN jet model.      \label{tab_model}} 	
 \begin{tabular}{cccccccccccc} 
 \hline 
 \hline 
 \noalign{\smallskip} 
 Model & $L_{\rm kin}/L_{\rm AGN}$  & $n_{\rm H}$ 	&	$F_{\rm cal}$	&	$B$ & $\eta_{\rm p}$	&  $\eta_{\rm e}$ & & & &	\\
               & 	        & 	 (cm$^{-3})$     & 	        & 	 (G)           & 	        & 	& &	    \\
\noalign{\smallskip} 
 \hline 
\noalign{\smallskip} 
W1	 & 3$\times$10$^{-3}$	&	10$^{4}$	&	1	      &	3$\times$10$^{-5}$	&	0.2	& 0.02 & &  \\

\noalign{\smallskip} 
W2	&3$\times$10$^{-3}$	&	10$^{4}$	&	 1 & 2$\times$10$^{-3}$	      &	 0.2	&	0.02 & &	 	\\

\noalign{\smallskip} 
W3	&  7$\times$10$^{-4}$	&	120	&	0.5        &	25$\times$10$^{-5}$	&	0.5	&  0.4 & &	  \\

\noalign{\smallskip} 
W4 &		3$\times$10$^{-3}$	&	10$^{4}$	& 1		&60$\times$10$^{-5}$        &	0.3	&	0.1  & & 	 	\\
\noalign{\smallskip} 
\hline 
 \hline 
 \noalign{\smallskip} 

Model	&	$\delta$  & $B$ & $R$	&	$T$	& $L_{\rm IR}$ 	& $r$  &  $n_1$ & $n_2$ & $\gamma_{\rm b}$ & $\gamma_{\rm max}$	\\

              &	  &  (G) & 	 (cm) &	(K)	&  (erg s$^{-1}$)	&  (cm) &   &  &  &  & \\
\noalign{\smallskip} 
 \hline 
\noalign{\smallskip} 
Jet	&	1.2  & $10^{-4}$ & 2$\times 10^{19}$  & 100 & 	1.5$\times 10^{42}$ &  2.2$\times 10^{20}$  & 2.2 & 3.3 & $10^{4}$ & $10^{7}$\\

  \noalign{\smallskip}
  \hline
  \end{tabular}
  \end{table*} 


\subsection{AGN wind model}

In  \cite{Lamastra16} we presented a physical model for  the  gamma-ray emission  produced by CR particles accelerated by the AGN-driven wind observed in the circumnuclear molecular disk of \textsc{NGC 1068} \citep{Krips11,Garcia14}. The gamma-ray emission has a hadronic component  that originates from the decay of neutral pions produced  in inelastic collisions between accelerated protons  and ambient protons,  and a leptonic component produced by IC scattering, and Bremsstrahlung radiation produced by the primary and secondary accelerated electrons.
In this paper we have updated the AGN wind model by implementing the effects of the interaction of gamma-ray photons with  the extragalactic background light (EBL), and the electromagnetic cascades initiated by high energy photons during their propagation in the intergalactic medium.\\

 We assumed  that  protons and electrons are accelerated  by the shocks produced by the AGN wind interaction with the surrounding ISM.  The structure of the AGN wind-shock system  is composed by a forward shock expanding into the ISM,  and a reverse shock moving inwards, which are separated by a contact discontinuity. Both the reverse and forward shock are likely sites of particle acceleration with an efficiency depending on  radiative losses  \citep{Nims15,Bustard17,Romero18}. 
We assumed that diffusive shock acceleration (DSA)  operates  in AGN-driven shocks.  Thus, the accelerated particle  energy distribution can be expressed as a power-law with spectral index $p\simeq$2 and  an exponential high-energy cut-off  \citep{Bell78,Bell78b,Blandford78,Drury83}, and it  is constrained by the wind dynamics and kinetic energy, and by the magnetic field in the shock region.  The wind dynamics and kinetic energy are  derived from observations in the millimetre band  \citep{Krips11,Garcia14}. The observationally derived quantities  are the outflowing gas mass  $M_{\rm out}$ $\simeq$1.8$\times$10$^{7}$ M$_{\odot}$,  the average radial extent of the outflow  $R_{\rm out}\simeq$100 pc, and  the projected radial outflow velocity  $v_{\rm out}\simeq$(100-200) km s$^{-1}$, which give a wind kinetic luminosity equal to  $L_{\rm kin}=(0.5-1.5)\times10^{42}$ erg s$^{-1}$.
 For the magnetic field  $B$ we assume the volume average ISM magnetic field strength $B=6\times(\Sigma_{\rm gas}/0.0025$ g cm$^{-2})^a \mu$G where $a\simeq$0.4-1  and $\Sigma_{\rm gas}$=(0.01-0.05) g cm$^{-2}$ is the disk gas surface density \citep{Robishaw08,Lacki10,Mcbride14}.  
 The latter   corresponds to a gas number density  $n_{\rm H}=$(115-460) cm$^{-3}$  assuming a  cylindrical geometry with radius of 350 pc and  vertical scale height $h\simeq$10 pc  \citep{Schinnerer00}. The  gas number density determines the efficiency of hadronic losses ($F_{\rm cal}$) and free-free losses. The  energy loss of relativistic electrons  by IC scattering is determined by  the AGN radiation energy density at the location of the shock $U_{\rm rad}=L_{\rm AGN}/4\pi cR_{\rm out}^2$,  we modelled the AGN SED with the synthetic spectrum  computed by \citep{Sazonov04}, and we normalized it to the bolometric luminosity of  \textsc{NGC 1068}  L$_{\rm AGN}$=(0.4-2.1$\times$10$^{45}$) erg s$^{-1}$ \citep{Bock00, Alonso11,Garcia14, Marinucci16}.

The particle energy-loss processes, and the size of the accelerator, determine the maximum energy of accelerated particles.  As discussed in \cite{Lamastra16},  the timescale on which particle are advected in the AGN wind ($\tau_{adv}=R_{\rm out}/v_{\rm out}$)  is larger than the the residence time of CR protons  and electrons in the central nuclear disk of the galaxy, while diffusion losses are not taken into account (see e.g. \cite{Peretti18} for a description of diffusion losses in the nuclei of starburst galaxies). The latter constitutes a source of uncertainty in the computation of the maximum proton energy that we estimate to be  $E_{\rm max} \simeq$10-100 TeV.

Although \textsc{NGC 1068} is a relatively nearby source, above few tens of TeV the effects of the interaction of gamma-ray photons with the IR emission spectrum of the EBL start to be important and determine the absorption of a substantial fraction of the flux, imprinting a well defined cut-off at the observed spectrum. The electron-positron pairs produced in this way, however, scatter off the photons of the CMB, triggering an electromagnetic cascade that reprocess the absorbed flux. 
Gamma-rays  may also be absorbed by radiation fields  within the  galaxy.  The MID-IR and FAR-IR thermal dust  emission from the $ \sim $1.5-2 kpc  starburst ring \citep{Spinoglio05} that surrounds the acceleration region could provide the target photons for pair production with TeV gamma-rays.
In this paper we take into account  also these physical processes in deriving the gamma-ray spectrum predicted by  the AGN wind model. 
For the EBL absorption  we adopt the model by \cite{Dominguez11}, and we find that it dominates over absorption of gamma-rays inside the galaxy. 
For the high energy electromagnetic cascades we use the convenient analytical treatment provided by \cite{Berezinsky16}.
The latter  demonstrate that -- in the case of nearby sources and monoenergetic injection at $E_{\rm inj}$-- a quite good approximation for the resulting cascade spectrum can be given in terms of  a broken power law. Specifically, the spectrum can be written as:
\begin{equation}
\label{reproc}
F(E) = \begin{cases}
k\left( \frac{E}{E_X}\right)^{-0.5}~, & E<E_X ~,\\[8pt]
k\left( \frac{E}{E_X}\right)^{-0.98}~, & E_X<E<E_{\rm inj}
\end{cases} 
\end{equation}
where
\begin{equation}
E_X=\frac{1}{3}\left( \frac{E_{\rm inj}}{m_ec^2}\right)^2\epsilon_{\rm CMB},
\end{equation}
and $\epsilon_{\rm CMB}=6.3\times 10^{-4}$ eV is the photon energy of background radiation.
The normalization constant $k$ can be derived by integrating the flux absorbed through the interaction with the EBL. This description is strictly valid for monoenergetic gamma-rays but for our purposes we can use it even if we have a relatively broad absorbed spectrum. In that case we can expect that the main effect is to have, instead of the pronounced breaks given by Eq. \ref{reproc}, smoother changes of the slopes. In our calculations we assume $E_{\rm inj}=$ 100 TeV.

\begin{figure*}
\begin{center}
\includegraphics[width=11 cm,trim={0cm 0cm 0cm 5cm},clip]{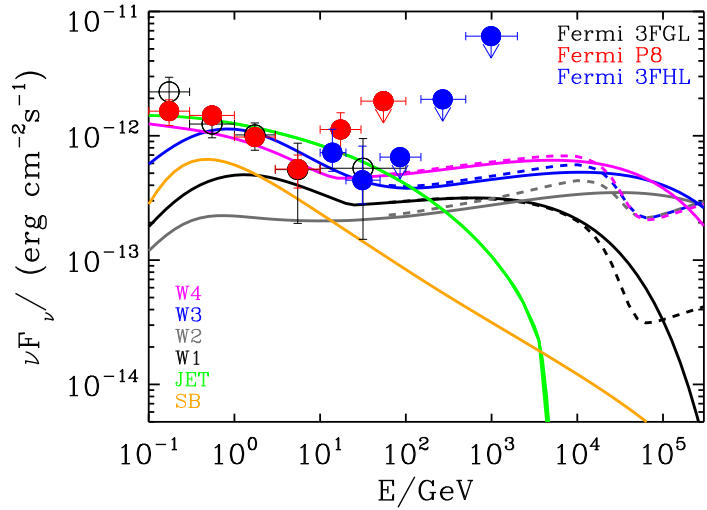}
\caption{Gamma-ray spectrum of \textsc{NGC 1068}.  The data points are from \cite{Acero15} (3FGL), from \cite{Lamastra16} (P8), and from \cite{Fermi_3FHL} (3FHL). The green 
 solid line shows the AGN jet  model prediction. The black, grey, blue, and magenta solid lines  show the AGN wind model predictions derived  by \cite{Lamastra16}, while dashed lines show the AGN wind model prediction taking into account the EBL absorption and electromagnetic cascades.  For comparison, the gamma-ray spectrum predicted by the starburst model \citep{Eichmann15} is shown with the orange solid line.}
\label{NGC1068_model}
\end{center} 
\end{figure*}

In Figure \ref{NGC1068_model} we compare the resulting gamma-ray spectra with those obtained by \cite{Lamastra16} for  different combinations of model parameters (see Table \ref{tab_model}).  In these models the galaxy and AGN properties  were varied within their observational ranges, and  both  standard electron ($\eta_e$) and proton ($\eta_p$)  acceleration efficiencies (W1 and W2 models), and acceleration efficiencies larger than those predicted by the standard   acceleration theory were adopted (W3 and W4 models). As expected, the spectra predicted in this work differ significantly from those obtained by \cite{Lamastra16} only above  $E$=10 TeV. As input of the simulations we consider only the revised spectral models derived in this work. 
The W1 and W2 models underestimate the {\it Fermi}-LAT spectrum, however, even if the gamma-ray emission predicted by these models is not the dominant contribution in the {\it Fermi}-LAT band,  it  could be relevant  in the CTA band, so we performed the simulations also for these models.

\subsection{AGN jet model}

An alternative to the hadronic scenario discussed above is a leptonic model based on the assumption that the observed gamma-ray emission is produced by relativistic electrons flowing into a jet. In fact, as proposed by \cite{Lenain10}, the large gamma-ray emission from  NGC 1068 could originate from the IC emission of relativistic electrons accelerated within a mildly relativistic collimated outflow expelled from the nucleus. In this scenario electrons could scatter both the synchrotron radiation (SSC) and the IR radiation of the surrounding environment. For reasonable values of the model parameters the latter component is expected to dominate.

We reproduce the inverse Compton spectrum of \cite{Lenain10} using the code fully described in \cite{Maraschi03}. Briefly, the emission is produced in a spherical region with radius $R=2\times 10^{19}$ cm carrying a tangled magnetic field with intensity $B=0.1$ mG in relativistic motion, parametrized by the relativistic Doppler factor $\delta=1.2$. Electrons -- following a smoothed broken power law energy distribution with normalization $K=12.5$ cm$^{-3}$ and slopes $n_1=2.2$ and $n_2=3.3$ below and above a Lorentz factor $\gamma_b=10^4$ and with maximum Lorentz factor $\gamma_{max}$=$10^7$ -- emit synchrotron and IC radiation.  Following \cite{Lenain10} we assume that the IR external field is modeled as a black body at a temperature $T=100$ K, a luminosity $L_{\rm IR}=1.5\times 10^{42}$ erg s$^{-1}$ occupying a spherical volume of radius $r=2.2\times 10^{20}$ cm. 

The AGN jet model parameters are summarized in Table  \ref{tab_model}, and  the IC spectrum is reported in Fig. \ref{NGC1068_model}  (green solid line). The {\it Fermi}-LAT data set the (relatively soft) slope of the high-energy electrons. The maximum energy of the electrons is a free parameter. We assume a maximum Lorentz factor $\gamma_{\rm max}=10^7$, i.e. a maximum energy of about 10 TeV.  However, for the large value assumed for the maximum Lorentz factor of the emitting electrons, the resulting IC spectrum is only moderately dependent on the precise value of $\gamma_{\rm max}$, since the cut-off visible in Fig. \ref{NGC1068_model} is shaped by the transition to the Klein-Nishina regime. Indeed, for larger value of $\gamma_{\rm max}$ the high-energy cut-off does not move to much larger energies.


\begin{figure*}
\begin{center}
\includegraphics[width=9 cm,angle=90]{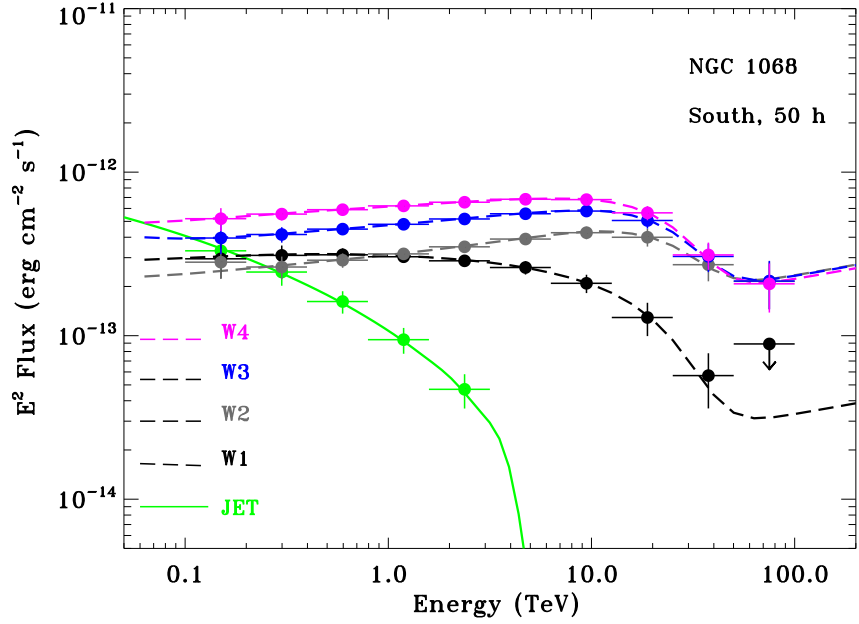}
\caption{Simulated spectra  of \textsc{NGC 1068}, color coded as in Fig.~\ref{NGC1068_model}. 
See text for details.}
\label{sim}
\end{center} 
\end{figure*}

 	 \section{Simulations} \label{lama_simulations}

%
           \begin{table*} 	
 \begin{center} 	
 \caption{Array of {\tt ctools} simulations.  W1$\div$W4= AGN wind models. Jet= AGN jet model. \label{lama:table:sims_BL}} 	
 \begin{tabular}{llllllcll} 
 \hline 
 \hline 
 \noalign{\smallskip} 
 Model	&	Site	&	zAngle	&	IRF	&	Expo	         &	Bins	& Energy	&	Number		\\
                & 	        & 	 (deg)       & 	        & 	  (h)           & 	        & (TeV)	&	                	 	    \\
 \noalign{\smallskip} 
 \hline 
 \noalign{\smallskip} 
W1	&	S	&	20	&	{\tt South\_z20\_average\_50h}	        &	50h	&	10	& 0.1--100 &	1000		\\
 \noalign{\smallskip} 
W2	&	S	&	20	&	{\tt South\_z20\_average\_50h}	        &	50h	&	10	& 0.1--100 &	1000			\\
 \noalign{\smallskip} 
W3	&	S	&	20	&	{\tt South\_z20\_average\_50h}	        &	50h	&	10	& 0.1--100 &	1000			\\
 \noalign{\smallskip} 
W4	&	S	&	20	&	{\tt South\_z20\_average\_50h}	        &	50h	&	10	& 0.1--100 &	1000			\\
 \noalign{\smallskip} 
Jet	&	N	&	20	&	{\tt North\_z20\_average\_50h}	        &	50h	&	5	& 0.1--3.2 &	1000		\\
Jet	&	S	&	20	&	{\tt South\_z20\_average\_50h}	        &	50h	&	5	& 0.1--3.2 &	1000		\\
  \noalign{\smallskip}
  \hline
  \end{tabular}
\end{center}
  \end{table*} 


%
%
%
%

\subsection{Set-up}
We performed our simulations using the analysis package {\tt ctools} v.1.4.2 \cite{Gammalib_ctools_2016}\footnote{\href{http://cta.irap.omp.eu/ctools/}{http://cta.irap.omp.eu/ctools/.} }, 
following the method described in \cite{Landoni19}, and the public CTA instrument response  files
(IRF, v.\ prod3b-v1). 
The  source is located at RA(J$2000)=40.669583$ deg,  Dec(J$2000)=-0.013333$ deg,
so it is visible from both CTA sites, at zenith angles of $\sim29^\circ$ from the North and $\sim24^\circ$ from the South. We therefore chose the {\tt North\_z20\_average\_50h} and {\tt South\_z20\_average\_50h}  IRFs. 
The CTA Southern site baseline array is located at the European Southern Observatory in the Atacama Desert (Chile), and  is planned to be  composed of 4 large-sized telescopes (LSTs), 25   medium-sized telescopes (MSTs), and 70 small-sized telescopes (SSTs), while in the  Northern site,  located at the Observatorio del Roque de los Muchachos on the island of La Palma (Spain), 4 LSTs and 15 MSTs  are planned to be  installed.

The SSTs  contribution to the CTA sensitivity becomes dominant over the MSTs above a few TeV. Since wind models predict emission up to energies of about 100~TeV, the northern site results less favourable when considering wind models with respect to the jet one, which, on the contrary, shows a possible fall-off below 10~TeV. For this reason we discuss our simulations performed for the southern site and for both CTA sites when considering wind and jet models, respectively. 
A summary of the inputs to our simulations is reported in Table~\ref{lama:table:sims_BL}. 

In the model definition XML file, the spectral model component was introduced as a 
``FileFunction'' type, 
with the differential flux values described according to 
\begin{equation}
M_{\rm spectral}(E)=N_0 \frac{dN}{dE},
\end{equation} 
where $N_0$ is the normalisation. 
%
We only considered the instrumental background included in the IRFs files 
({\tt CTAIrfBackground}) and no further contaminating astrophysical sources in the 
5\,deg field of view (FOV) we adopted for event extraction.

%
We considered a set of 10 energy bins covering an energy range reported in 
Table~\ref{lama:table:sims_BL} (Col.\ 7). For the last model, we only 
considered 5 bins. 
In each bin,  {\tt ctobssim}  was used to create event lists based on our input models. 
%
Then, the task {\tt ctlike} was used to  fit each spectral bin with a power-law model 
\begin{equation}
M_{\rm spectral}(E)=k_0 \left( \frac{E}{E_0} \right) ^{\Gamma},
\end{equation}
where $k_0$ is the normalisation (or {\tt Prefactor}, in units of ph\,cm$^{-2}$\,s$^{-1}$\,MeV$^{-1}$), 
$E_0$ is the pivot energy ({\tt PivotEnergy} in MeV), 
and  $\Gamma$ is the power-law photon index ({\tt Index}). In the fits 
{\tt PivotEnergy} was fixed at $10^{6}$\,MeV, while {\tt Prefactor} and {\tt Index} were left to vary. 
We thus obtained the spectral parameters and fluxes 
of our gamma ray source in each bin by using maximum likelihood model fitting. 
Statistical uncertainties of the parameters were also calculated, 
as well as the test statistic (TS) value \citep[][]{Cash1979,Mattox1996:Egret}. 
For each spectral model $N=1000$  sets of 
statistically independent realisations were created 
by adopting a different seed for the randomization
in order to reduce the impact of variations between individual realisations 
\citep[see, e.g.][]{Gammalib_ctools_2016}. 
We thus obtained a set of 1000 values of each spectral parameter (and TS)
from which 1000 values of fluxes were calculated in each energy bin. 
For each energy bin, we adopted as flux and error the 
mean and the square root of the standard deviation 
obtained from the distribution of fluxes\footnote{Mean 
flux $\overline{F_{\rm sim}}  = \frac{1}{N}\sum_{k=1}^{N}F_{\rm sim}(k)$, 
standard deviation $s^2_{\rm sim}=\frac{1}{N-1}\sum_{k=1}^{N}(F_{\rm sim}(k)-\overline{F_{\rm sim}})^2$. }.

\subsection{Results}

Figure \ref{sim} shows the spectra obtained with the simulations. For each model we report the input spectrum and the reconstructed spectrum in bins of energy. 
In the case of the AGN jet model, the simulation shows that the spectrum can be detected only up to energies $E\simeq$3 TeV. In fact,  the high-energy cut-off in the spectrum caused by the maximum energy of accelerated leptons  implies no signal at the highest energy. 
On the other hand, the hadronic component predicted by the AGN wind model produces a spectrum that extends up to  $E\simeq$100 TeV. The simulations show that the  high energy emission 
 can be tracked in the whole energy range up to $E\simeq$100 TeV for the W2, W3, and W4 models, and up to $E\simeq$50 TeV  in the case of W1 model.
 
\begin{figure*}
\begin{center}
\includegraphics[width=10 cm,angle=0,trim={0cm 12cm 0cm 2cm},clip]{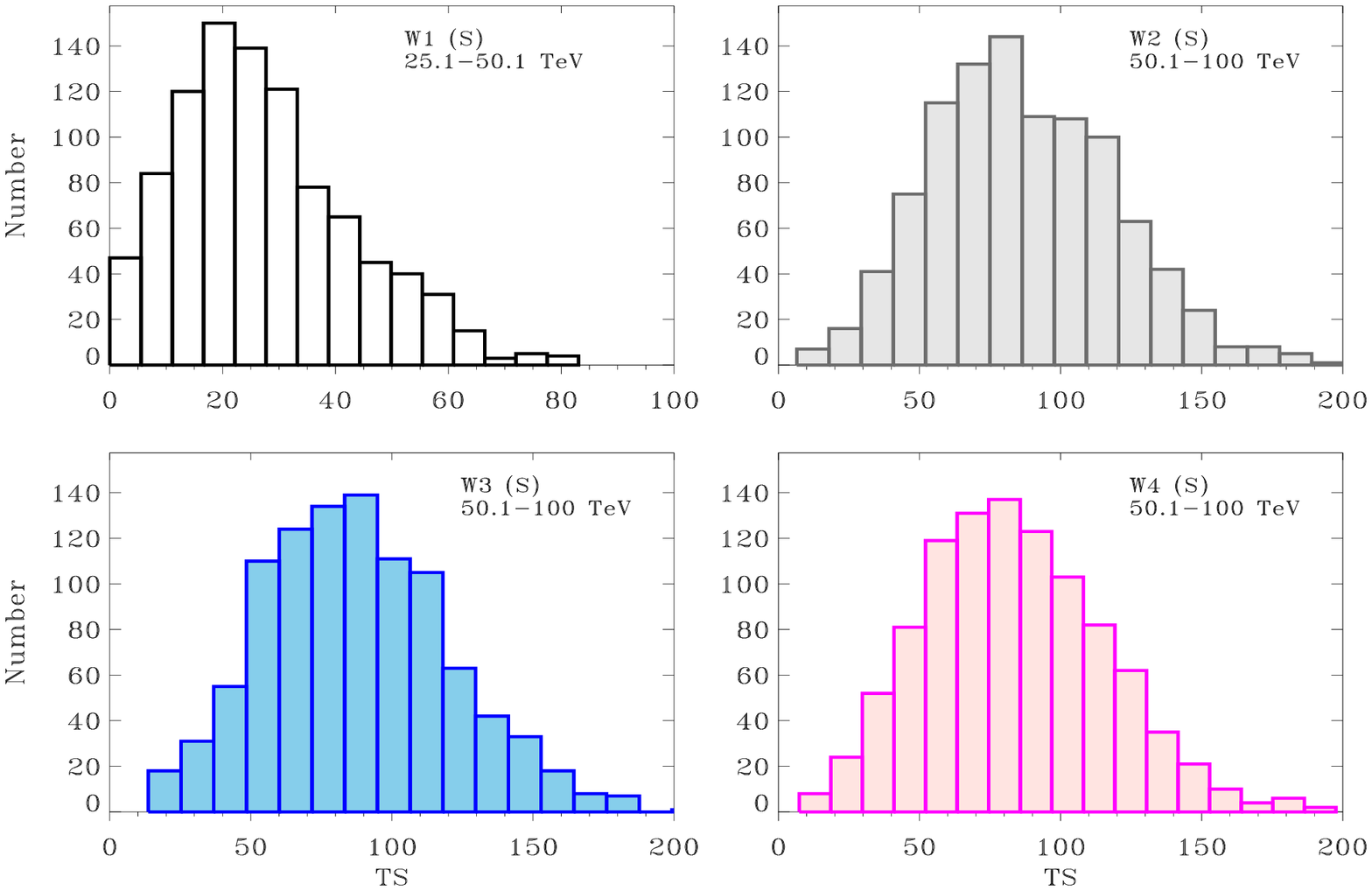}
\caption{TS value distribution of the 1000 realizations  in the last bin where we obtained a significant detection: 25.1-50.1 TeV (W1), and 50.1-100 TeV (W2,W3,W4) for the AGN wind model.}
\label{significance}
\end{center} 
\end{figure*}

 These simulations imply that,  considering 50 hours of observations,   CTA can be effectively used to constrain the two different emission models. In fact, a detection at $E\gtrsim$10 TeV will provide a smoking gun for the existence of a hadronic component in the gamma-ray spectrum of \textsc{NGC 1068}.
To give an idea of  the significance of the detection, we report the average and the standard deviation of the TS value obtained from the $N$=1000 sets of statistically independent  realisations,  
 in the last bin where we obtained a significant detection: 12.6-25.1 TeV in the case of W1 model: $\langle TS \rangle_{\rm W1}$=111.3$\pm$33.1, and for the energy bin 25.1-50.1 TeV for the other wind models: $\langle TS \rangle_{\rm W2}$=194.2$\pm$48.4, $\langle TS \rangle_{\rm W3}$=225.4$\pm$52.5, $\langle TS \rangle_{\rm W4}$=233.1$\pm$52.0.  The only marginal critical point is the last detected energy bin for which we show in Figure \ref{significance} the whole TS distributions.  Figure \ref{significance}  shows that for the majority  of the realizations the TS value is greater than 25 implying a  significant detection also at the highest energies.

Moreover, the expectedly unprecedented sensitivity of CTA will allow us to reach a photon statistics able  to determine fundamental features of the gamma-ray spectrum, like  the position of the  high energy cut-off. Therefore, as it will be discussed in the next Section, CTA  observations of  \textsc{NGC 1068} will provide valuable constraints on the  physics of the shock waves produced by AGN-driven winds.

\section{Discussion}\label{discussion}

The simulations presented in this work indicate that with 50 hours of observation with CTA we will be able to constrain  emission models in \textsc{NGC 1068}. In fact, in case of signal detection at energies $E\gtrsim$10 TeV, a pure jet model will be firmly excluded. 
The  simulations indicate that the  sensitivity and energy coverage of CTA allows us to measure with  good accuracy  the normalization and the position of the high energy cut-off of the gamma-ray spectrum.

Extending the spectrum to TeV energies is crucial to better constrain the non-thermal high energy luminosity of the source, and to compare it with the AGN and galaxy bolometric luminosity. From this comparison we might be able to determine if the hadronic gamma-ray emission is dominated by star formation or nuclear activity. At the highest energies, the AGN wind model predicts a a spectral hardening above $\sim$50 TeV. With a longer exposure CTA might reveal this spectral feature providing stringent constraints on hadronic models.

The detection of a high energy cut-off  provide information on the maximum energy of accelerated particles which depends on shock parameters. In the case of protons, the maximum energy depends on the shock velocity,  size of the accelerator, and on the magnetic field strength  in the shock region. It has been suggested that  protons accelerated by AGN-driven winds reach energies up to $E_{\rm max}\simeq$10-100 PeV \citep{Tamborra14,Wang16_neutrini,Lamastra16,Lamastra17,Liu18}, the observational evidence of this hypothesis could have important implications for models of PeV neutrinos sources detected by IceCube \citep{Aartsen14,Aartsen15}.

The comparison between the total non-thermal luminosity  of \textsc{NGC 1068}   with the AGN wind kinetic luminosity  inferred from millimetre spectroscopy (see Section \ref{AGN_wind}) could  constrain particle acceleration efficiencies ($\eta_{\rm p}$ and $\eta_{\rm e}$) and the calorimetric efficiency ($F_{\rm cal}$) in AGN-driven winds.
The W1 and W2 models proposed by \cite{Lamastra16}  are obtained  by viewing AGN-driven wind as  SN-driven wind analogue, thus values of $\eta_{\rm p}$ and $\eta_{\rm e}$ determined from SN observation in the radio and gamma-ray band are adopted  \citep[and references therein]{Keshet03,Thompson06,Tatischeff08,Lacki10}. However, the parameters that determine the physics of shock waves in the region near the AGN, like the  pre-shock magnetic field and ISM density, could be not the same as those in non-active galaxies resulting in  larger acceleration efficiencies than those predicted by the standard acceleration theory  (W3 and W4 models). 
The determination of the normalization of the spectrum at TeV energies puts strong constraints on the values of $\eta_{\rm p}$ and $\eta_{\rm e}$ in AGN-driven winds.

As for the calorimetric fraction, to date there is no measurement of $F_{\rm cal}$ in AGN-driven winds. Assuming  the standard SN paradigm for the gamma-ray emission in \textsc{NGC 1068},  values of 
$F_{\rm cal}\gtrsim$1 have been obtained by various authors \citep{Ackermann12,Wojaczynski17,Wang18}. 
This argues for a significant contribution of the active nucleus to the gamma-ray emission.  Also in the other two Seyfert galaxies detected in the gamma-ray band,   high efficiencies of gamma-ray production are predicted by  the starburst scenario:  $F_{\rm cal}\simeq$1 for \textsc{NGC 4945}, and  $F_{\rm cal}\gg$1 for Circinus \citep{Lenain10,Hayashida13,Wojaczynski17,Wang18}.  
Similarly to \textsc{NGC 1068}, AGN-driven molecular winds and weak radio jets  have been  observed in \textsc{NGC 4945} and  Circinus \citep{Zschaechner16,Henkel18,Elmouttie98,Lenc09}. These galaxies also show  similarity of their Eddington-scaled X-ray and gamma-ray luminosity with $L_{0.1-100  \rm GeV}/L_{\rm Edd}\simeq$10$^{-4}$ and $L_{2-10 \rm keV}/L_{\rm Edd}\simeq$10$^{-2}$ in all three sources \citep{Wojaczynski15,Wojaczynski17}, corroborating a correlation between the AGN activity  and gamma-ray emission. 

\begin{figure*}
\begin{center}
\includegraphics[width=13 cm,angle=0]{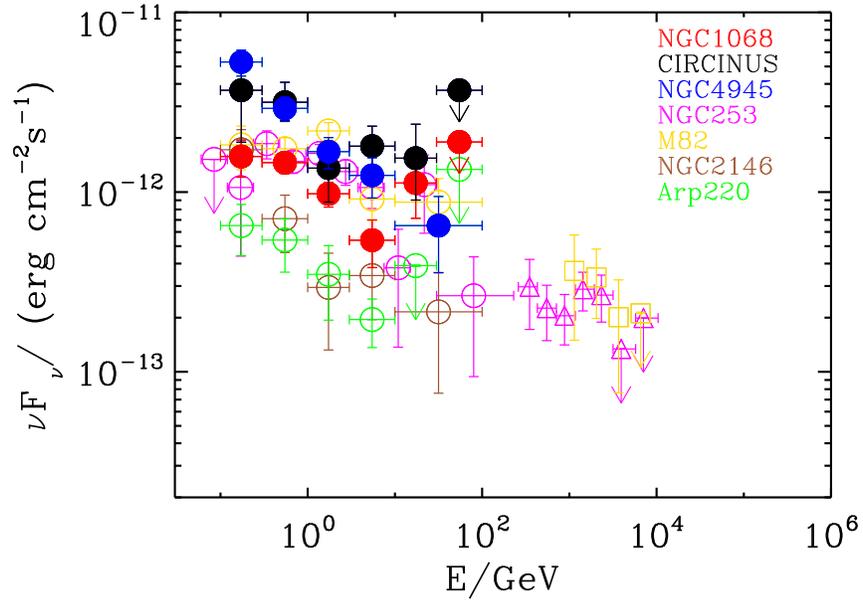}
\caption{Gamma-ray spectra of Seyfert galaxies (filled symbols) and starburst galaxies (open symbols) detected by \textit{Fermi}-LAT and IACTs. The data points are from \cite{Ackermann12, Acero15, Lamastra16,Fermi_3FHL,Wojaczynski17,Hayashida13,Acciari09,Tang14,Peng16,NGC253_HESS_18}.}
\label{starburst}
\end{center} 
\end{figure*}

The gamma-ray spectra of these Seyfert galaxies, as well as of  starburst galaxies, detected by \textit{Fermi}-LAT and current  IACTs, are shown in Figure  \ref{starburst}. 
The starburst galaxies \textsc{M82} and \textsc{NGC 253}  have already been detected in the VHE band by  VERITAS, and H.E.S.S., respectively \citep{Acciari09,NGC253_HESS_12,NGC253_HESS_18}.  The nearby ultra luminous infrared galaxy Arp 220 was observed by  MAGIC and VERITAS  but no significant excess over the background was found \citep{Albert07,Fleischhack15}.

The all-sky coverage and good sensitivity of the CTA full array will provide an unprecedented  opportunity to improve our understanding of the gamma-ray emission in  Seyfert and starburst galaxies that is crucial to test models of gamma-ray  and neutrino sources that contribute to  the extragalactic backgrounds observed by \textit{Fermi}-LAT and IceCube \citep{Ackermann15,Aartsen14,Aartsen15}.

        \section*{Acknowledgements}


%
The authors acknowledge contribution from the grant INAF CTA--SKA,
``Probing particle acceleration and $\gamma$-ray propagation with CTA and its precursors'' (PI F.\ Tavecchio). \\
AL acknowledges contribution from the grant INAF CTA--SKA ``ASTRI/CTA Data Challenge'' (PI P. Caraveo). \\
%
This research has made use of the NASA/IPAC Extragalactic Database (NED) which is operated
by the Jet Propulsion Laboratory, California Institute of Technology, under contract with the
National Aeronautics and Space Administration. \\
%
This research made use of {\tt ctools}, a community-developed analysis package for Imaging Air Cherenkov Telescope data.
ctools is based on {\tt GammaLib}, a community-developed toolbox for the high-level analysis of astronomical gamma-ray data. \\
%
This research has made use of the CTA instrument response functions provided by the CTA Consortium and Observatory,
see \href{https://www.cta-observatory.org/science/cta-performance/}{https://www.cta-observatory.org/science/cta-performance/} 
(version prod3b-v1) for more details. \\
%
          We gratefully acknowledge financial support from the agencies and organizations
          listed here:
\href{http://www.cta-observatory.org/consortium_acknowledgments}{http://www.cta-observatory.org/consortium\_acknowledgments}.\\
%
This paper went through internal review by the CTA Consortium.\\
We thank  L. Stawarz  and J. P. \ Lenain as internal CTA reviewers.

%

%





 \bibliographystyle{elsarticle-num} 
 \bibliography{biblio_1068_CTA.bib}



\end{document}